\documentclass[twocolumn,prl]{revtex4-1}

\usepackage{graphicx}
\usepackage{amsmath}
\usepackage{amssymb}
\usepackage{wasysym}

\newcommand{\bra}[1]{\langle #1|}
\newcommand{\ket}[1]{|#1\rangle}
\newcommand{\expectation}[1]{\langle #1\rangle}
\newcommand{\inner}[2]{\langle #1 | #2 \rangle}

\begin{document}

\title{Localisation in space and time in disordered-lattice open quantum dynamics}

\author{Sam~Genway}\affiliation{School of Physics and Astronomy, The
  University of Nottingham, Nottingham NG7 2RD, United Kingdom}

\author{Igor~Lesanovsky}\affiliation{School of Physics and Astronomy, The
  University of Nottingham, Nottingham NG7 2RD, United Kingdom}

\author{Juan~P.~Garrahan}\affiliation{School of Physics and Astronomy, The
  University of Nottingham, Nottingham NG7 2RD, United Kingdom}

\date{\today}

\begin{abstract}
We study a two-dimensional tight-binding lattice for excitons with on-site disorder, coupled to a thermal environment at infinite temperature.  The disorder acts to localise an exciton spatially, while the environment generates dynamics which enable exploration of the lattice.   Although the steady state of the system is trivially uniform, we observe a rich dynamics and uncover a dynamical phase transition in the space of temporal trajectories.  This transition is identified as a localisation in the dynamics generated by the bath.  We explore spatial features in the dynamics and employ a generalisation of the inverse participation ratio to deduce an ergodic timescale for the lattice.
\end{abstract}

\maketitle

Probing the dynamics of quantum systems out of equilibrium is a big challenge of current research in physics.  As well as being of fundamental interest, 
a particular application is the study of exciton transport, relevant to materials ranging from thin-film dyes~\cite{Tennakone2013} and conjugated polymers~\cite{Bolinger2011,Bardeen2011} to semiconductor nanostructures~\cite{Wheeler2013, Scholes2006}.  Exciton transport is also of great importance in light harvesting materials~\cite{Cheng2009, Scholes2000, Yang2002} such as the Fenna-Matthews-Olson complex~\cite{Fenna1974}.   Of particular interest is the interplay between disorder, which leads to exciton-localisation effects, and dissipation which facilitates exciton transport~\cite{Nejad2011,Nejad2013,Vlaming2013}.  A complete understanding of such systems is still being sought~\cite{Xiong2012} and this motivates the exploration of the rich dynamical features which emerge generically in dissipative disordered systems.

In this work, we seek to understand general features in the dynamics of an exciton in a large disordered lattice coupled to an infinite-temperature thermal environment.  While disorder acts to localise excitons spatially~\cite{Anderson1958, Cutler1969, Lee1985}, the environment generates dynamics which allow the entire lattice to be explored.  The dynamical phenomena which arise in such systems is studied using a ``thermodynamics of trajectories'' formalism~\cite{Ruelle2004,Garrahan2007,Lecomte2007,Merolle2005,*Baule2008,*Gorissen2009,*Jack2010,*Giardina2011,*Nemoto2011,*Chetrite2013}.  Using this method, we will show that while the steady state of the model at infinite temperature is trivial, with all regions equally likely to be occupied, the dynamics show complex features including a dynamical phase transition in the space of trajectories.   

\begin{figure}[tb]
\includegraphics[width=8.57cm]{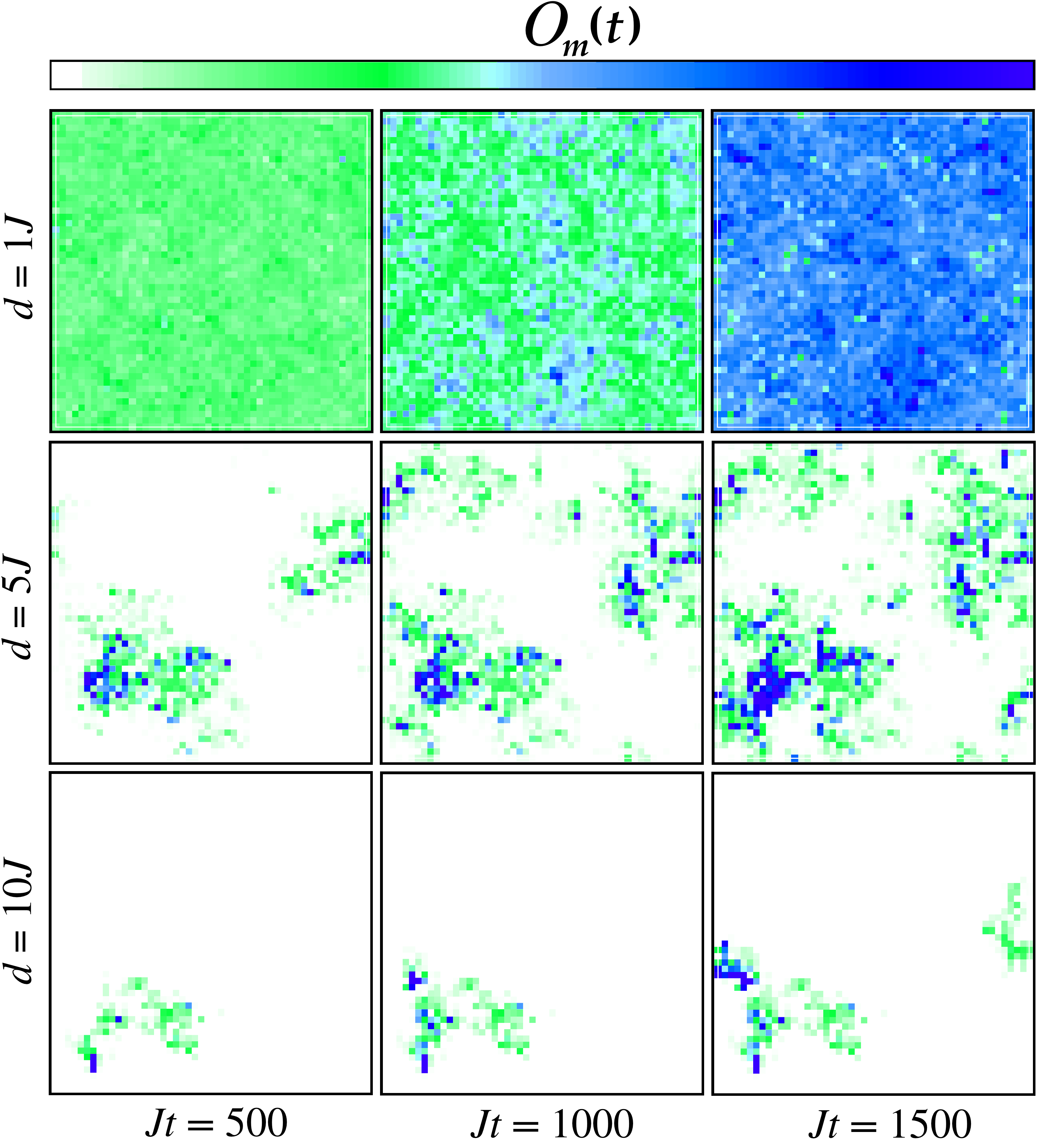} 
\caption{(Colour online.) Excitonic occupation of regions of an $N = n\times n$ disordered lattice, with $N=10^4$ sites, coupled to an infinite temperature bath.  Shown is the lattice-site occupation $O_m(t)$ for different times $t$ (left to right) and different disorder strengths $d$ (top to bottom).  See main text for details.}
\label{fig1}
\end{figure}

The transition takes the form of a localisation transition in time: there is an \emph{inactive} phase, where the exciton remains localised in a particular state, and an \emph{active} state, where the environment induces a rapid change between states and the exciton explores all space.   In Fig.~\ref{fig1} we show this effect becomes more pronounced as the strength of the disorder is increased.  Such active-inactive trajectory transitions are characteristic of glasses and other classical and quantum systems with pronounced dynamical metastability \cite{Garrahan2007,Hedges2009,*Speck2012,*Speck2012b,Garrahan2010,Garrahan2011,*Ates2012}. Our findings suggest that a general feature of the dynamics in disordered systems coupled to environments is the existence of an increasingly super-Poissonian temporal distribution for the jumps between lattice sites as disorder is increased.  Remarkably, this dynamical behaviour exists even at infinite temperature.

We study a two-dimensional tight-binding model with on-site energies chosen randomly from a Gaussian distribution.  We are interested in the parameter space in which the disorder is sufficiently strong such that all eigenstates are localised within the size of the lattice.  Specifically, we consider a square lattice with $N=n\times n$ sites and periodic boundary conditions with Hamiltonian
\begin{equation}
H = \sum_m \varepsilon_m \ket{m}\bra{m} + J\sum_{\expectation{mm'}} \ket{m}\bra{m'} = \sum_i E_i \ket{i}\bra{i}\,.
\end{equation}
Each state $\ket{m}$ has a wavefunction centred on a site with label $m$ and corresponding energy $\varepsilon_m$ drawn randomly from a Gaussian distribution, with variance ${d^2}$ and zero mean.  The size of $d$ will set the disorder strength.   The site index $m$ is related to the coordinates $(x,y)$ of the lattice site via $m=x+n(y-1)$, with $1\le  m \le N$.  In the second term, $\expectation{mm'}$ denotes a sum over nearest neighbours and we will choose units for energy such that the hopping integral $J$ equals unity.  We will use indices $i$ and $j$ for eigenstates of $H$, where $H\ket{i} = E_i \ket{i}$.

The effect of dissipation is introduced by coupling the system to a bath of harmonic modes with Hamiltonian
\begin{equation}
H_b = \sum_k \omega_k b_k^\dag b_k\,.
\end{equation}
These couple to the system via the coupling Hamiltonian
\begin{equation}
H_{sb} = S \otimes B = \sum_m c_m \ket{m}\bra{m} \otimes \sum_k h_k (b_k+b_k^\dag) \,,
\end{equation}
where the parameters $c_m$ are also selected randomly from a Gaussian distribution with zero mean and a variance we will specify.  Under standard manipulations (Born, Markov and secular approximations), we find a master equation diagonal in the basis of eigenstates $\dot{P}_i = (\mathbb{W})_{ij} P_j$, where $P_i$ is the occupation probability of the eigenstate $\ket{i}$.  The master operator $\mathbb{W}$ has elements $(\mathbb{W})_{ij}$ given by
\begin{equation}
(\mathbb{W})_{ij} =  W_{j\rightarrow i}  - r_i\delta_{i,j}
\label{eq:W}
\end{equation}
where the transition rates $W_{j\rightarrow i}$ are given by
\begin{equation}
W_{j\rightarrow i} =  J(\omega_{ji})\,\, |\bra{j}S\ket{i}|^2\,.
\label{eq:Wij}
\end{equation}
and $J(\omega_{ji}) = 2\pi \sum_k |h_k|^2 \delta(\omega_k - \omega_{ji})$ is the spectral density of the bath 
with $\omega_{ji} = E_j - E_i$.  We will study the case of a bath with temperature $T=\infty$ such that the rates satisfy $W_{i\rightarrow j} = W_{j\rightarrow i}$.  In this work, we consider an Ohmic bath with $J(\omega) = \omega$;  this choice fixes the variance of the parameters $c_m$.

At long enough times, we anticipate that all knowledge of the initial location of the exciton will be lost and the probability of finding the exciton anywhere in the lattice will be uniform in accordance with the $T=\infty$ distribution.  To ascertain how long the exciton has spent in different regions of the lattice we integrate the eigenstate occupation probabilities $P_i(t)$ over time and define $O_i(t) = \int_0^t dt' P_i(t')$.  We express these occupation times in the local basis as $O_m(t) = \int_0^t dt' \sum_i |\inner{m}{i}|^2 P_i(t')$ 
\begin{figure}[htb]
\includegraphics[width=8.57cm]{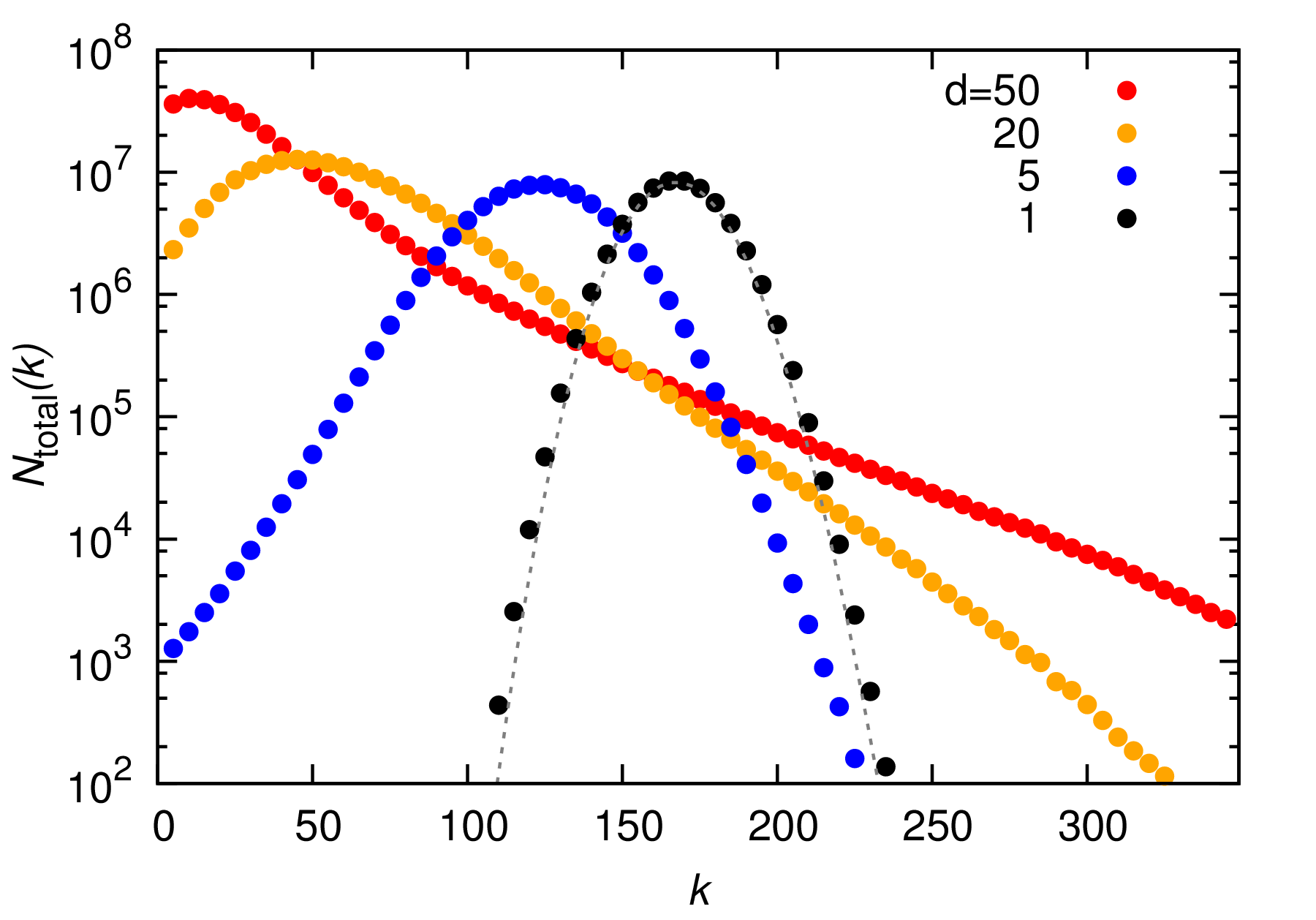}
\caption{(Colour online.)  Histograms of the number of jumps $k$ in time intervals $t=300/J$ for simulations (as in Fig.~\ref{fig1}) with $10^8$ jumps in total.  Plotted is the number of time intervals in which $k$ jumps occur, $N_\text{total}(k)$, for different strengths of disorder $d$ (labelled).  Shown (dashed line) is a fit to the $d=1$ points assuming a Poisson distribution.}
\label{fig2}
\end{figure}
Plotted in Fig.~\ref{fig1} are snapshots at different times of three trajectories at different disorder strengths $d$, all prepared in the same local initial state.  At small $d$ it is clear that the exciton moves almost uniformly in space and time, with the lattice having been occupied uniformly after short times.  Conversely, as $d$ is increased we find the exploration of the lattice becomes far from uniform in time, with large dwell times in certain regions and quick jumps between other regions.  This effect becomes increasingly pronounced as $d$ is increased and it will be studied in greater detail later in the paper.   

First we look at the statistics for jumps between states, captured by the probability $\pi_t(K)$ that there are $K$ jumps between states in a time $t$.  Shown in Fig.~\ref{fig2} are histograms reflecting this distribution for long time intervals $t=300/J$ at different values of $d$.   While at small $d=1$, the distribution is close to Poissonian, as $d$ is increased, it becomes progressively broad, indicating that \emph{rare trajectories}, where $K$ is much smaller or larger than the mean, becoming increasingly likely.

We note that at long times the probability distribution takes the large-deviation form~\cite{Touchette2009}
$\pi_t(K) \simeq e^{-t\varphi(K/t)}$  where $\varphi(k)$ is a large-deviation function of the average jump rate, or \emph{activity}, $k=K/t$.   The associated moment generating function is also of large-deviation form $Z_t(s) = \sum_K \pi_t(K) e^{-sK} \simeq e^{t\theta(s)}$, with $s$ a conjugate field to the number of jumps $K$.  The function  $\theta(s)$ is analogous to (minus) a free energy for trajectories, with discontinuities in the derivatives of $\theta(s)$ corresponding to dynamical (or trajectory) phase transitions~\cite{Garrahan2007,Lecomte2007,Garrahan2010}.  The activity $k_s = -\partial_s\theta(s)$ will be used as an order parameter, with $k_{s=0}=k$ the average jump rate of the physical problem  (\emph{i.e.} with no $s$ field applied).  We can find $\theta(s)$ as largest eigenvalue~\cite{Lecomte2007} of the modified master operator $\mathbb{W}_s$, described by
\begin{equation}
(\mathbb{W}_s)_{ij} =  e^{-s} W_{j\rightarrow i}  - r_i\delta_{i,j}\,.
\label{eq:Ws}
\end{equation}
This operator generates the dynamics of $s$-biased ensembles of trajectories via $\partial_t P_i(s) = \sum_j(\mathbb{W}_s)_{ij} P_j(s)$.  The eigenstate of $\mathbb{W}_s$ corresponding to the largest eigenvalue $\theta(s)$ gives the occupation probabilities $P_i(s)$ (of level $i$) associated with the trajectories that dominate at a certain $s$.  When $s=0$ this is the stationary state which, since we are studying the case of an infinite temperature environment, has equal probability for all levels, such that $P_i(s=0) = 1/N$ for all $i$.  At $s\ne 0$,  $P_i(s)$ indicate the occupations for rare trajectories which are more ($s<0$) or less ($s>0$) \emph{active} than those of the average dynamics.

\begin{figure}[htb]
\includegraphics[width=8.57cm]{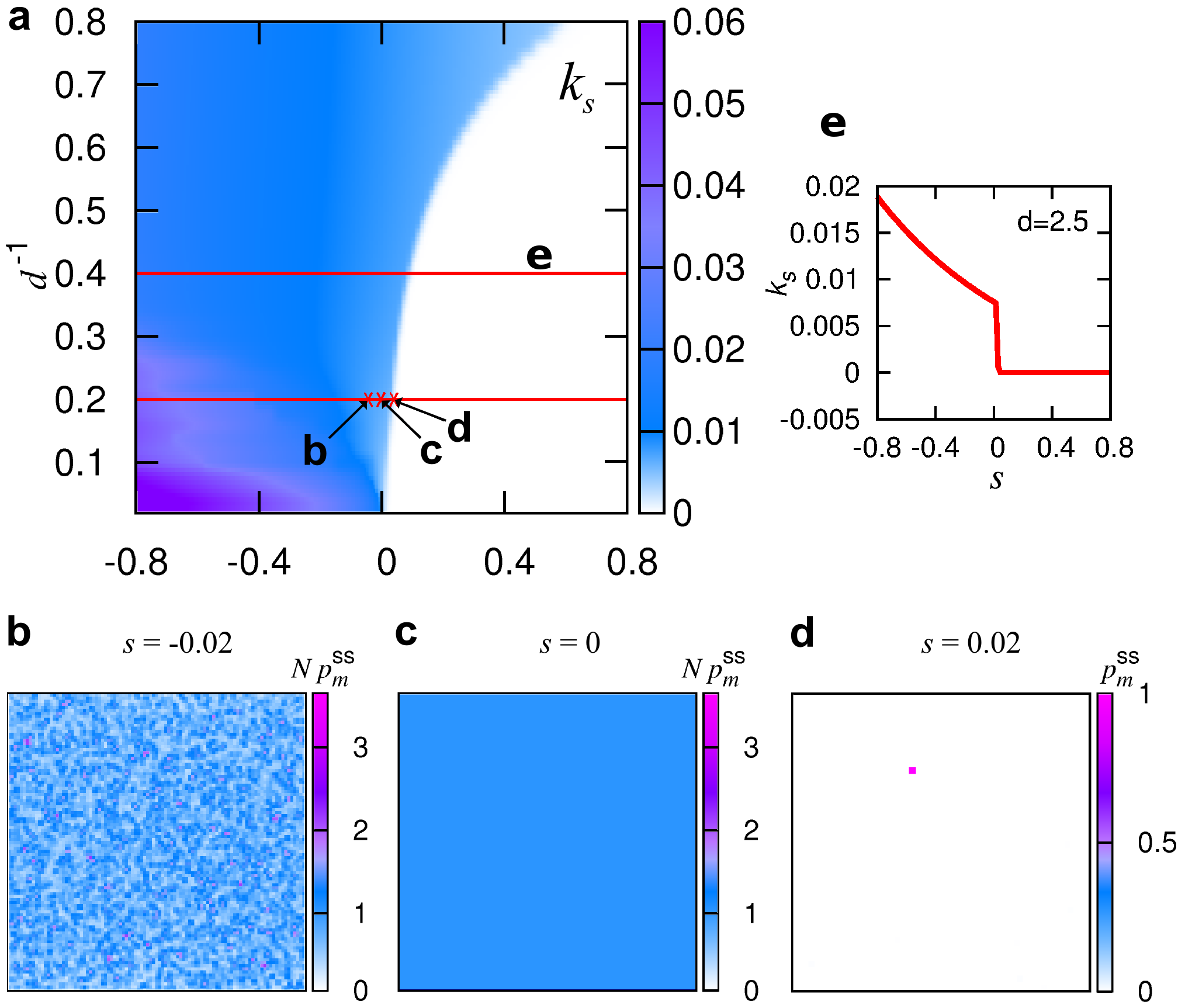} 
\caption{(Colour online.) (a) Dynamical phase diagram for the $N=100\times 100$ disordered lattice in Fig.~\ref{fig1} as a function of inverse disorder strength and thermodynamic field $s$ (see main text).  (b,c,d) Steady-state lattice-site occupations in the long-time limit,  $P_m(s)$.  Shown are the occupation probabilities for the parameter values as labelled on (a).  (e) A slice through the dynamical phase diagram with parameter values labelled on (a). }
\label{fig3}
\end{figure}

We now apply this thermodynamic approach to the dynamics in disordered exciton-lattices.  We obtain the dynamical phase diagram for the model~\eqref{eq:Wij} by plotting the activity $k_s$ as a function of the inverse disorder strength $d^{-1}$ and the $s$-field.  The data, found from exact diagonalisation, are shown in Fig.~\ref{fig3}(a).  We find a first-order phase boundary for $s>0$, which appears to tend to $s=0$ in the limit of infinite disorder strength.  We will show that existence of such a transition in a single-particle system is due to states becoming increasingly disconnected at higher disorder.   The is existence of a transition is consistent with the increasingly long tails on $\pi_t(K)$ as $d$ is increased, shown in Fig.~\ref{fig2}.  The large-$s$ phase has $k_s=0$ where, for these inactive rare trajectories, the exciton does not jump between eigenstates.  Therefore, the exciton must remain localised.  This result we confirm in Fig.~\ref{fig3}(d), where we see that the $s$-biased steady-state occupation probability is zero everywhere apart from close to a particular site.  In contrast, Fig.~\ref{fig3}(c) shows the uniform ($T=\infty$) distribution across the lattice at $s=0$.  This distribution is only slightly perturbed from uniform if $s$ is decreased further to $s=-0.02$, as demonstrated in Fig.~\ref{fig3}(b).

\begin{figure}[htb]
\includegraphics[width=8.57cm]{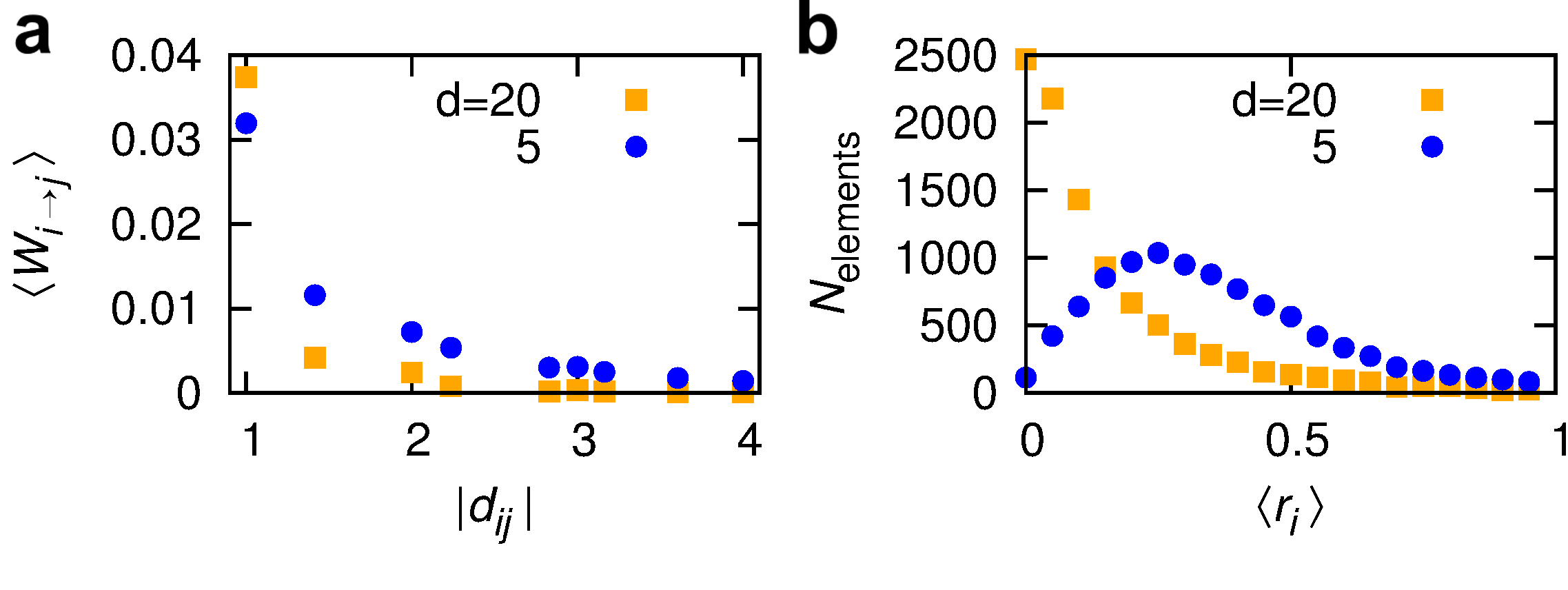} 
\caption{(Colour online.)  Statistics of matrix elements of the master operator $\mathbb{W}$ for single realisations with $10^4$ sites.  (a) The average rate of transitions between states $\ket{i}$ and $\ket{j}$ centred on lattice sites a distance of $d_{ij}$ lattice periods apart for two disorder strengths $d=5$ and 20.  (b) Histograms of the number, $N_\text{elements}$, of diagonal elements $r_i$ in $\mathbb{W}$ with different magnitudes.  While the shape of the distribution depends strongly upon $d$, the variance of the elements only changes from 0.274 to 0.229 as $d$ is increased from $d=5$ to $d=20$.}
\label{fig4}
\end{figure}

The transition in the dynamics can be understood as a localization transition in the master operator~\eqref{eq:Ws}.  The transition matrix $\mathbb{W}_s$ of this $T=\infty$ model is Hermitian such that we can draw analogy with a quantum Hamiltonian:  at large $d$, where the eigenstates are tightly localised, $\mathbb{W}_s$ is equivalent to the sum of a ``hopping'' part $e^{-s} W_{j\rightarrow i}$ which determines the jumps between localised states, and ``on-site energies'' $r_i$.  $\mathbb{W}_s$ exhibits a localisation transition which is crossed by tuning $s$.  We find the existence of the transition is not associated with finite-size effects, but with the presence of long-range hopping terms $W_{j\rightarrow i}$ which favour a delocalised state~\cite{Rodriguez2003,Metz2013,Biddle2011}.  In Fig.~\ref{fig4}(a) we explore the effect of $d$ on the log-range hopping terms $W_{j\rightarrow i}$.  Plotting the mean values of these matrix elements between states centred on sites separated by $d_{ij}$ lattice spacings, we find that hopping terms beyond nearest-neighbour sites are significantly larger at smaller $d$.  In Fig.~\ref{fig4}(b) we plot the distribution of effective on-site terms for the dynamics, $r_i$.  Interesingly, this effective on-site disorder has a distribution which changes shape with $d$, but the variance of the elements $r_i$ changes little with $d$.  Thus we infer that the transition is controlled by the effective hopping terms, with the transition line approaching $s=0$ at large $d$ due to the increased size of long-range hopping matrix elements.  When crossing the transition by tuning the $s$ field, the size of the effective hopping terms in comparison to the effective disorder is decreased as $s$ is increased, such that the localised phase is reached.

We now explore the behaviour of the system in both space and time for different values of the disorder strength $d$.  We showed in Fig.~\ref{fig1} that as the exciton jumps through the lattice, it moves around some regions very quickly and dwells for long times in other regions.  In the limit of long times, the exciton will have occupied all regions of the lattice for equal fractions of the evolution time, in accordance with the infinite-temperature thermal distribution.  We wish study the time scale for the exciton to explore the lattice and consider how the number of persistent sites not visited by the exciton decreases over time at different values of $d$.  For this, we introduce a generalised inverse participation ratio (GIPR) $p_t$ at finite times $t$ defined by
\begin{equation}
p_t = \frac{1}{t}\sum_m O^2_t(m) 
\end{equation}
If a single lattice site $m$ is occupied initially, we have $p_{0}=1$.  More generally, for an initial state $\ket{\psi}$, $p_{0}$ is determined by overlaps of $\ket{m}$ with the initial state wavefunction and is given by the conventional inverse participation ratio in the basis of position states~\cite{Gogolin2011,*Genway2012a}, $\sum_m |\inner{m}{\psi}|^4$.  
At times long enough for the entire lattice to be explored, $p_t$ approaches $p_\infty = 1/N$, reflecting the thermal distribution.  This sets ergodic timescale as it requires the exciton to explore the entire lattice.  

\begin{figure}[htb]
\includegraphics[width=8.57cm]{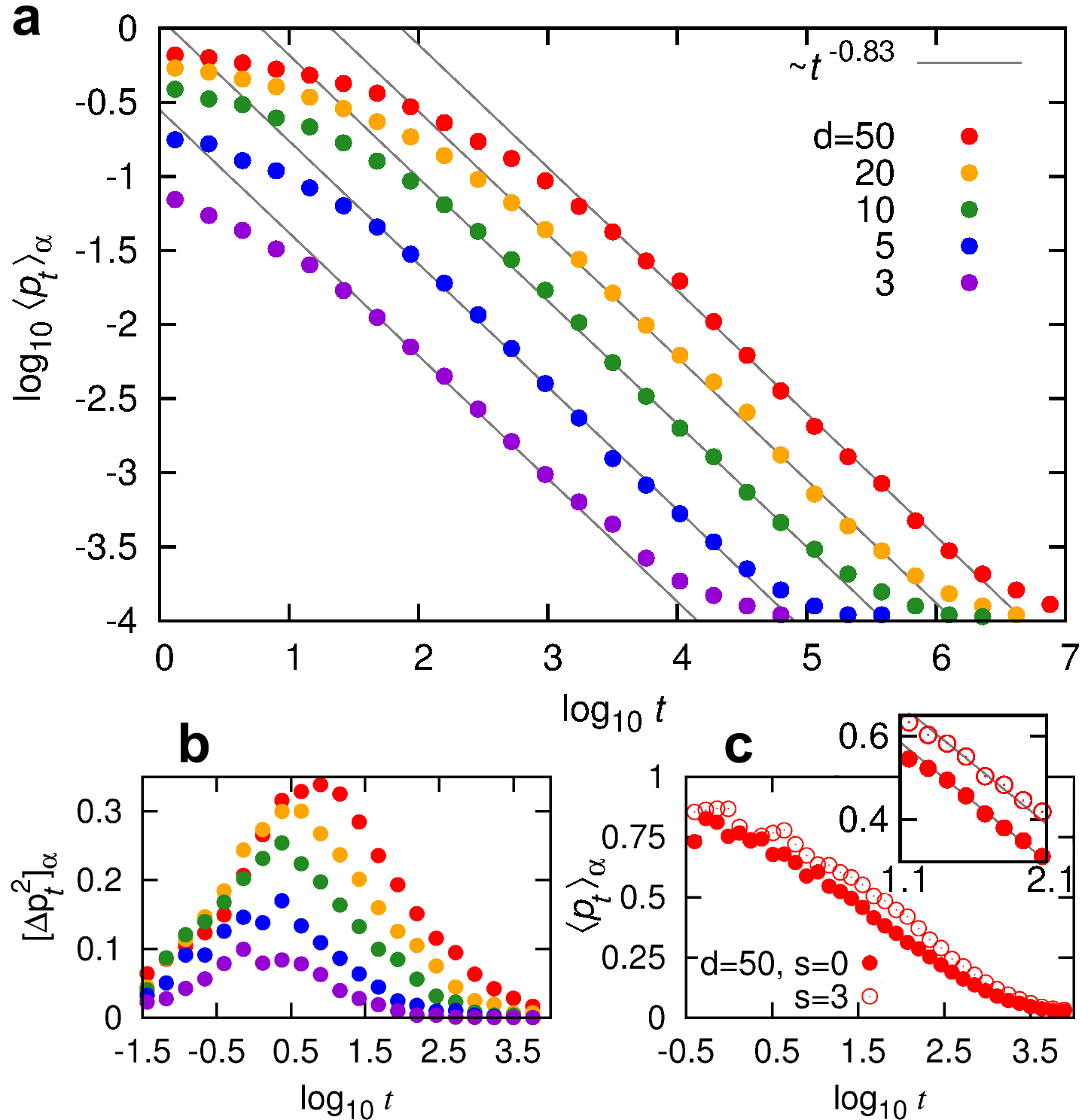} 
\caption{(Colour online.) GIPRs for the simulations in Fig.~\ref{fig1}.  (a)  The GIPR averaged over trajectories $\expectation{p_t}_\alpha$.  The results are an average of $10^3$ simulations of $10^3$ jumps and 5 trajectories of $10^9$ jumps, to capture accurately the GIPR average across different initial states as well as the long-time, initial-state-independent behaviour.  Initial states are selected randomly from the uniform distribution.  Shown are different disorder strengths $d$, as labelled. (b)  The variance of GIPRs for different trajectories $[\Delta p_t^2]_\alpha$ as a function of time for the same disorder strengths as in (a).  (c)  GIPRs averaged over trajectories for $d=50$ with different initial states.   Shown is a comparison of initial states drawn from the uniform distribution and initial states drawn from the inactive-state distribution with occupation probabilities $P_i(s)$ for $s=3$ (as labelled). Shown \emph{inset} is a magnified region with linear fits.  These demonstrate that the curves are separated in time by a factor $10^{0.315}\simeq 2.1$.}
\label{fig5}
\end{figure}

To study the statistics associated with trajectories, we now introduce the label $\alpha$ which identifies a trajectory.   We estimate the GIPR from simulated trajectories using Eq.~\eqref{eq:W}.  Fig.~\ref{fig5}(a) illustrates, as a function of time, an average of the GIPR over trajectories, $\expectation{p_t}_\alpha$, for different disorder strengths $d$.  In Fig.~\ref{fig5}(b), we show the variance of the GIPR over trajectories, $[\Delta p_t^2]_\alpha = \expectation{p_t^2}_\alpha-\expectation{p_t}^2_\alpha$, for the same values of $d$.  We see that for $d=20$, the ergodic timescale is around two orders magnitude longer than at $d=3$.  This can be contrasted with the corresponding average jump rate (see Fig.~\ref{fig2}) which only decreases by a factor of $\apprle 4$.  
The decay of the GIPR takes an approximate power-law form $p_t \sim t^{-\eta}$ for time scales greater than the mean jump rate $k$ but smaller than the ergodic time.  From our numerics, we find $\eta \simeq 0.83$ across the range of $d$ we consider in Fig.~\ref{fig5}.

As $d$ is increased, the states where the exciton dwells for a long time become increasingly inactive.  This situation is analogous to finite-temperature diffusion of a classical particle in a random potential with a set of deep minima where the particle can get stuck for long times.  We expect that as $d$ is increased, the initial state will have a larger effect on the decay of $p_t$, with a slower initial decay if the initial state is inactive.  This is confirmed by our data for the GIPR by considering the variation between trajectories and the effect of choosing an inactive initial state.  The variance between trajectories $[\Delta p_t^2]_\alpha$ is shown in Fig.~\ref{fig5}(b).  We find $[\Delta p_t^2]_\alpha$ increases with $d$, as does the length of time to reach the peak in this variance, consistent with the existence of increasingly inactive regions on the lattice.  In Fig.~\ref{fig5}(c), for $d=50$ we compare the decay of the GIPR for initial eigenstates selected from $P_i(s=3)$  with those selected from the thermal distribution.  We find that the initial decay of the GIPR decays on a time scale approximately twice as long if we prepare the exciton in a state from the inactive dynamical phase.

In summary, we have explored the dynamics of excitons in disordered lattices when coupled to infinite-temperature Markovian baths.  While the steady-state occupation probability is simply the uniform distribution, we find rich features in the dynamics generated by the coupling to the thermal environment.  We have studied the counting statistics of jumps between states and we find, as the strength of the disorder is increased, the distribution of jumps deviates significantly from the Poissonian form found for weak disorder.  This deviation can be understood to relate to a dynamical phase transition in the space of temporal trajectories.  Upon increasing the disorder, the dynamical phase boundary imposes a greater effect on the physical dynamics of the system, with the exciton spending increasing periods of time in inactive regions where jumps between states are rare.  We also find that dynamical metastability also manifests in spatially heterogeneous dynamics, something which is very prevalent in glass forming systems \cite{Biroli2013}.  This effect was characterised with a GIPR, which captures the lifetime of persistent sites not visited by the exciton.  We find the GIPR decays in time as a fixed power-law for all disorder strengths $d$.  Of course, what we presented here is nothing else than a single particle problem where all non-trivial features are a consequence of the imposed disorder.  However, we may speculate that the combination that we find of metastability, a first-order transition in ensembles of dynamical trajectories, and dynamical heterogeneity will also be present in interacting systems which exhibit many-body localisation.

\begin{acknowledgments}
We wish to thank Hoda Hossein-Nejad for discussions.
We acknowledge support from the The Leverhulme Trust under grant No.~F/00114/BG.  The research leading to these results has received funding from the European Research Council
under the European Union's Seventh Framework Programme (FP/2007-2013) / ERC Grant
Agreement n.~335266 (ESCQUMA).
\end{acknowledgments}

\bibliographystyle{apsrev4-1}
\bibliography{anderson,dicke}

\end{document}